\documentclass [a4paper,12pt,oneside,final]{article}
\usepackage[left=26mm,top=25mm,right=26mm,bottom=15mm]{geometry}
\usepackage{graphicx} 
\usepackage{mathtools}

\usepackage{amsmath}
\usepackage{float}
\usepackage[dvipsnames]{xcolor}
\newcommand{\inlineeqnum}{\refstepcounter{equation}~~\mbox{(\theequation)}}
\usepackage{algorithm}
\usepackage[noend]{algpseudocode}

\makeatletter
\def\BState{\State\hskip-\ALG@thistlm}
\makeatother

\title{Asynchronous Merkle Trees}
\author{Anoushk Kharangate}
\date{November 2023}

\begin{document}

\maketitle

\begin{abstract}
\font\tenrm = cmr10 at 10pt
\tenrm
    Merkle trees have become a widely successful cryptographic data structure. Enabling a vast variety of applications from checking for inconsistencies in databases like Dynamo to essential tools like Git to large scale distributed systems like Bitcoin and other blockchains. There have also been various versions of Merkle trees like Jellyfish Merkle Trees and Sparse Merkle Trees designed for different applications. However, one key drawback of all these Merkle trees is that with a large data set the cost of computing the tree increases significantly, moreover insert operations on a single leaf require re-building the entire tree. For certain use cases building the tree this way is acceptable, however in environments where compute time needs to be as low as possible and where data is processed in parallel, we are presented with a need for asynchronous computation. This paper proposes a solution where given a batch of data that has to be processed concurrently, a Merkle Tree can be constructed from the batch asynchronously without needing to recalculate the tree for every insert.
\end{abstract}
\pagebreak

\begin{section}{Introduction}
Merkle trees, first introduced by Ralph Merkle\cite{Ralph Merkle} have been widely used in various applications, however the focus of this paper shall be on it's use in blockchains, specifically ones with a parallel execution environment. The specific use case depends on the design of each blockchain; for example, Bitcoin creates a Merkle tree of all the transactions and adds the hash in the root to the blockheader for SPV \cite{SPV}. Regardless of the different use cases, the core concept remains the same. It provides a cryptographic method to authenticate certain data maintaining consistency across nodes and providing log(n) proof to verify the same.

However there are a few notable drawbacks with the data structure that still plague these systems. This paper addresses the time and compute cost to not only construct the entire tree synchronously but also updating it sync. Specifically, a Merkle tree when constructed with N leaves needs to be completely recomputed if item N + 1 has to be appended. This introduces challenges in environments that are time bound and also parallel.
\end{section}
\begin{section}{Application in Blockchain Trees}
    As mentioned in the introduction Merkle trees are essential to the functioning of blockchains like Bitcoin. More recently so since the development of more feature rich and high performance systems like Solana \cite{Solana} and Monad \cite{Monad} which add concurrent execution to the mix, traditional Merkle trees become a bottleneck in terms of reaching consensus.
    Adding async trees to the consensus protocol can allow building trees of entire batches of transactions and accounts without blocking consensus. The tree can be built as each batch is being processed in it's own thread. This could fundamentally improve performance of the consensus protocol without sacrificing on essential features dependent on these trees.
\end{section}
\begin{section}{An Asynchronous Solution}
To address this I propose an Asynchronous Merkle Tree(AMT), a modified Merkle tree with the following key features: 

\begin{itemize}
    \item \textbf{Node Types}: The tree consists of 3 special nodes, \textit{Digest Node}, \textit{Layer Checkpoint Node(LC)} and \textit{Compound Node}.
     \begin{itemize}
          \item Two LC nodes combine to form a compound node.
          \item A LC node and a compound node form another compound node.
          \item A Digest node simply acts as a placeholder for a node from another batch.
     \end{itemize}
    \item \textbf{Ordering}: Each special node must contain an order bit indicating if it was a right child or a left child to it's parent.
    \item \textbf{Segregation}: Every node must have a batch bit indicating which batch it belonged to.
\end{itemize}
\subsection{Prerequisites}
    It is important to understand the the prerequisites for an AMT to work. An AMT should be based upon a Merkle Tree T that can be defined by the following requirements:
    \begin{itemize}
        \item {T has a height(h) where $h = log_2(n) \inlineeqnum\label{eqn:height}$ and n being the number of leaves.
        }
        \item T must have a maximum of M nodes, where M =  $\sum_{n}^{h} \frac{n}{2}$.
        \item Leaf nodes are the nodes $N_i, \{ i \in N \, | \, 0 \leq i \leq 2^D\} \inlineeqnum\label{eqn:leaf-nodes}$
    \end{itemize}
    
    We extend these to create an AMT with the following requirements:
    \begin{itemize}
        \item The leaves that are appended asynchronously are called Batches(B), the quantity of these batches and their leaves must be known in advance.
        \item The tree has to be initialised with as many \textit{Digest Nodes} ($D_i$) that act as placeholders for the nodes from other batches.
    \end{itemize}
        
\begin{section}{Async Appends}

  \begin{figure}[H]
            \centering
            \includegraphics[width=12 cm]{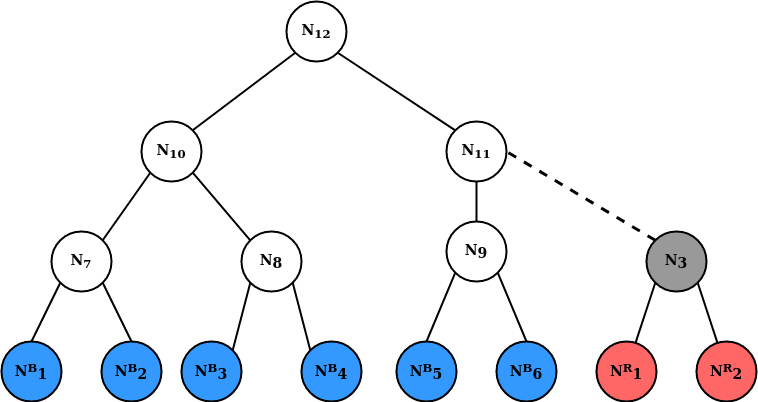}
           
    \caption[.]{
   A Merkle tree with depth 3. The {\color{NavyBlue} blue} batch consisting of nodes \{$N_1^{B}..N_6^{B}$\} and the {\color{BrickRed} red} batch consisting of the nodes \{$N_1^{R}..N_{14}^{R}$\}.}
            \label{fig:one}
         \end{figure}
    Suppose a list of data is being processed concurrently and a Merkle root has to be calculated with the output of the processed data as leaves of the tree. Let us assume the {\color{NavyBlue} blue} batch is processed first therefore we can only construct a tree of the length of the blue batch.
    We know the path for $N_5$ and $N_6$ goes through $N_{9}$ and $N_{11}$ as shown below.
    
    \begin{equation}\label{eq1}
      \begin{gathered}
         N_{9} = hash(N_5,N_6) \\
        N_{11} = hash(N_{9}) \\
        N_{12} = hash(N_{10},N_{11})
      \end{gathered}
    \end{equation} However, when the {\color{BrickRed} red} batch is processed some of it's leaves become a dependency for intermediary nodes as shown in Figure 1. $N_3$ should be a child to $N_{11}$ as shown below.
     \begin{equation}\label{eq2}
      \begin{gathered}
         N_{9} = hash(N_5,N_6) \\
        N_{11} = hash(N_{9}, N_3) \\
        N_{12} = hash(N_{10},N_{11})
      \end{gathered}
    \end{equation}

If we want to append these batches of leaves async, we need to first build a tree with the information we have. Since we know the total number of batches and their leaves in advance we know how many leaves we don't have, for those we can replace them with \textit{Digest Nodes}. Thus constructing a tree using the already processed batch and placeholders for the to be processed batches. This is important because it will help us understand which nodes become an inter-batch dependency from which batches.

     \begin{figure}[H]
            \centering
            \includegraphics[width=16 cm]{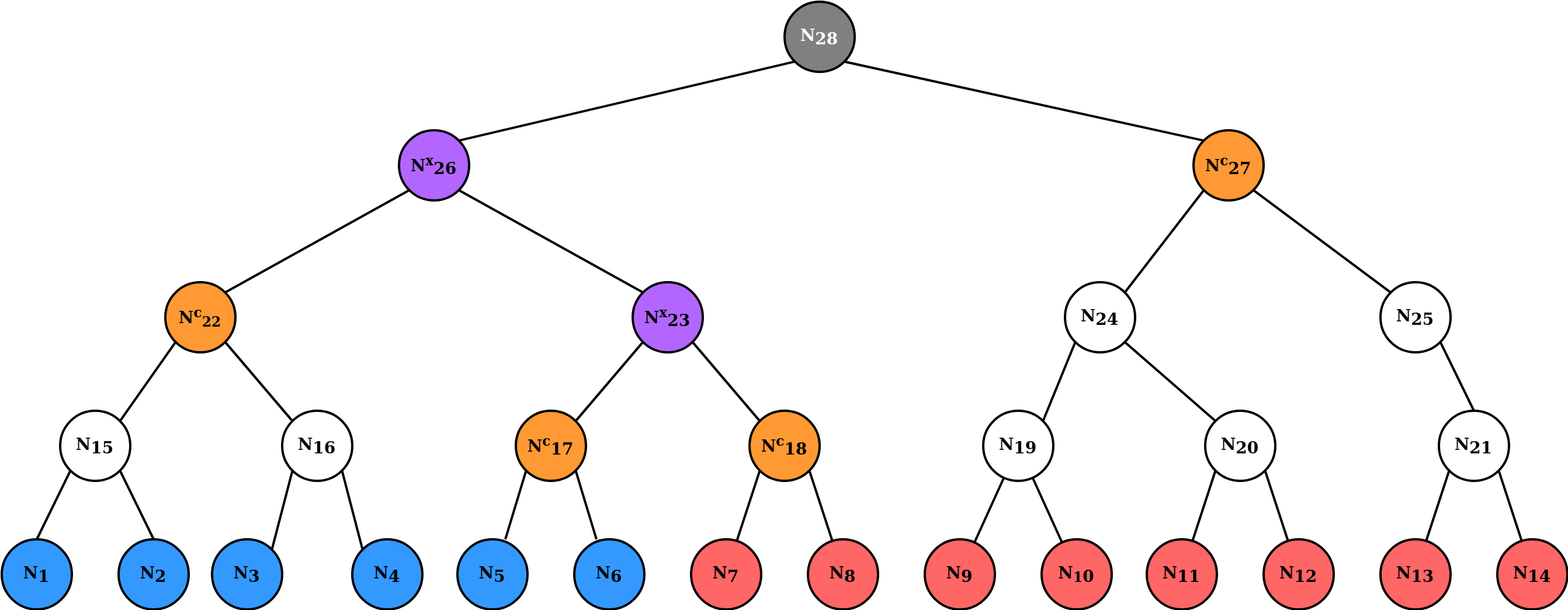}
           
    \caption[.]{
   An AMT with depth 4, with two batches. The {\color{NavyBlue} blue} batch consisting of nodes \{$N_1..N_6$\} and the {\color{BrickRed} red} batch consisting of the nodes \{$N_7..N_{14}$\}.}

            \label{fig:two}
         \end{figure}
As the tree is being constructed, if two nodes are being hashed together and belong to separate batches they create a \textit{Compound Node}. This means that those two nodes become \textit{Layer Checkpoint} Nodes i.e the maximum we can compute before we need to get the data from the other batch. These nodes can be stored in a array for when we want to finally commit the tree once all batches have arrived after processing and sync compute the root from the pre-computed nodes.

Taking the example from Figure 2, $N^{C}_{17}$ and $N^{C}_{18}$ belong to two separate batches can both can be pre-computed independently. However, when we want to compute $N^{X}_{23}$ which needs the above mentioned nodes it creates an inter-batch dependency. Therefore we store $N^{C}_{17}$ and $N^{C}_{18}$ in an array for later. This continues for $N^{X}_{26}$ and  $N_{28}$ as well since they both depend on a \textit{Compound Node} $N^{X}_{23}$ and  $N^{X}_{26}$ respectively, making $N^{C}_{22}$ and $N^{C}_{27}$ also part of the checkpoints.

However, in practice since we don't have the leaves from the {\color{BrickRed} red} batch yet, we need to make a new tree per batch with placeholders to calculate the checkpoints, as shown in Figure 3.
\begin{figure}[H]
            \centering
            \includegraphics[width=14 cm]{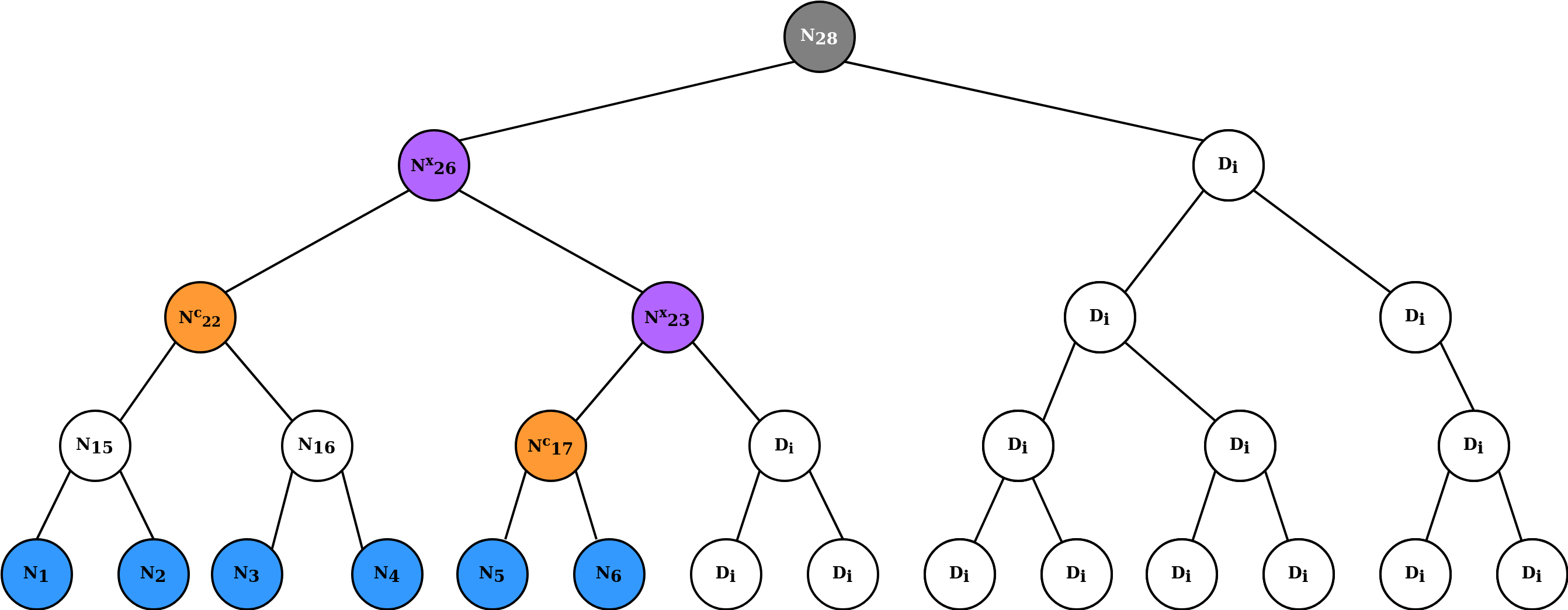}
           
    \caption[.]{
   An AMT with one batch and Digest Nodes. The {\color{NavyBlue} blue} batch consisting of nodes \{$N_1..N_6$\} and representing the {\color{BrickRed} red} batch as placeholders are the $D_i$ nodes.}
   
            \label{fig:three}
         \end{figure}

 \subsection{Node Structure}
 Each node must comprise of the following data:
 \begin{itemize}
     \item \textbf{Batch}: An integer indicating which batch the node belongs to.
       \begin{itemize}
           \item For compound nodes this can be set to a special integer that doesn't conflict with an existing batch id to signify that the node doesn't belong to one specific batch.
       \end{itemize}
     \item \textbf{Order}: An integer, either one or zero indicating left or right node.
     \item \textbf{Data}: The actual hash of the data.
\end{itemize}
\end{section}
\end{section}
\begin{section}{Sync Commit}
    Finally, after we have async appended all the batches and we want to compute the root we can commit the tree. Even though we are still computing the root synchronously, we only have to compute 3 nodes including the root node as compared to 14 nodes from the example in Figure 2. This significantly reduces the wait time to calculate the root and avoids blocking consensus for long.
     \begin{figure}[H]
            \centering
            \includegraphics[width=12 cm]{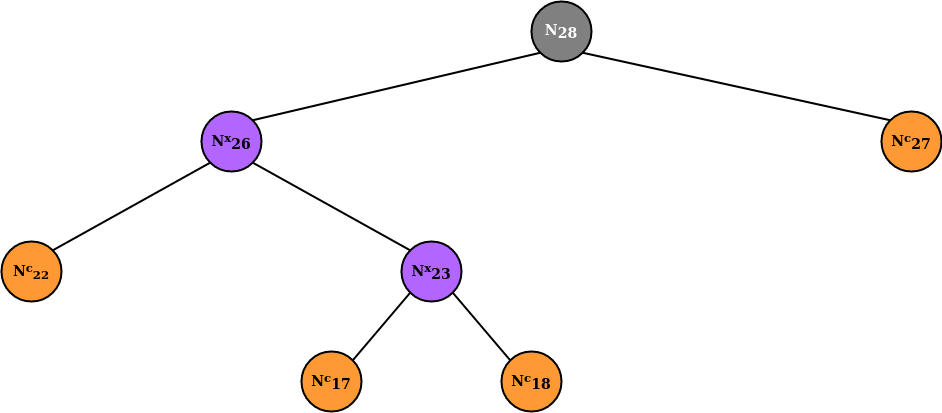}
           
    \caption[.]{
  The{ \color{orange} orange} indicates the LC Nodes that are stored, the nodes in {\color{Mulberry} pink} are sync computed when committing}
   
            \label{fig:four}
         \end{figure}
Iterating over the layer checkpoints, we can hash the first node with it's consecutive node and then hash the output with the next one until we end up with a single hash i.e the root. Since the trees should be deterministic we need to also read the order of the nodes before hashing them and then hash them in the sequence.

\end{section}
\begin{section}{Conclusion}
    To conclude we have created a new kind of Merkle tree called Asynchronous Merkle Tree(AMT) for blockchains with parallel virtual machines that can save on compute time by building the tree as each parallel batch is being processed. However, one trade off here is that since we need to make a different tree for every batch it could increase the memory footprint if the quantity of the batches increase significantly.
\end{section}

\end{document}